\documentclass[aps,pra,reprint,nofootinbib,twocolumn,superscriptaddress,showpacs,showkeys,longbibliography]{revtex4-1}

\usepackage{eurosym}
\usepackage{amsmath,amssymb,amstext}
\usepackage{ulem}
\usepackage[usenames,dvipsnames]{color}
\usepackage{graphicx}
\usepackage{braket}
\usepackage{natbib}
\usepackage{comment}
\usepackage{dcolumn}
\usepackage[english]{babel}
\usepackage{wasysym}
\usepackage{bm}
\usepackage[colorlinks,bookmarks=false,citecolor=blue,linkcolor=red,urlcolor=blue]{hyperref}

\usepackage{appendix}
\usepackage{CJK}

\begin{document}
\title{Effective two-level models for highly efficient inner-state enantio-separation based on cyclic three-level systems of chiral molecules}
\author{Chong Ye}
\affiliation{Beijing Computational Science Research Center, Beijing 100193, China}
\author{Quansheng Zhang}
\affiliation{Beijing Computational Science Research Center, Beijing 100193, China}
\author{Yu-Yuan Chen}
\affiliation{Beijing Computational Science Research Center, Beijing 100193, China}
\author{Yong Li}\email{liyong@csrc.ac.cn}
\affiliation{Beijing Computational Science Research Center, Beijing 100193, China}
\affiliation{Synergetic Innovation Center for Quantum Effects and Applications, Hunan Normal University, Changsha 410081, China}

\begin{abstract}
  Based on cyclic three-level systems of chiral molecules, we propose two methods to realize highly efficient inner-state enantio-separations of a chiral mixture with the two enantiomers initially prepared in their ground states. Our methods work in the region where the evolutions of the two enantiomers can be described by their corresponding effective two-level models, simultaneously. The approximately $100\%$-efficiency inner-state enantio-separations can be realized {when the probability  occupying the ground state of one enantiomer becomes $0$ by experiencing half-integer periods of its corresponding on-resonance Rabi oscillation} and in the meanwhile the other one still stays  approximately in the ground state, under the conditions that the two enantiomers are governed by the effective on-resonance and large-detuning two-level models, respectively. Alternatively, the exactly $100\%$-efficiency inner-state enantio-separation can be obtained when {the probabilities occupying the ground states of the two enantiomers} simultaneously experience half-integer and integer periods of their corresponding on-resonance and {detuned} (instead of largely-detuned) Rabi oscillations with final $0$ and $1$ probabilities occupying the ground state, respectively.
\end{abstract}
\date{\today}
\maketitle
\section{Introduction}
Chirality is important in chemistry, biotechnologies, and pharmaceutics due to the fact that the vast majority of chemical~\cite{A1}, biological~\cite{A2,A3,A4}, and pharmaceutical~\cite{A5,A7,A8,A6} processes are chirality-dependent. Yet, enantio-discrimination~\cite{PC0} and enantio-separation~\cite{PC1,PC2,PC3,PC4,PC5,Book1} of a chiral mixture are among the most important and difficult tasks in chemistry.
Some enantio-discrimination~\cite{CA1,CA2,CA3} and enantio-separation methods~\cite{PRL.121.173002,1808.08642} had been proposed based on the interferences between the electric- and magnetic-dipole transition moments. Since the magnetic-dipole
transition moments are usually very weak, alternative methods for enantio-discrimination~\cite{Hirota,Nature.497.475,PRL.111.023008,PCCP.16.11114,
ACI,JCP.142.214201,JPCL.6.196,JPCL.7.341,Angew.Chem.10.1002,KK} and  enantio-separation~\cite{PRL.87.183002,PRL.90.033001,PRA.77.015403,
PRL.99.130403,JCP.132.194315,JPB.43.185402,PRL.122.173202,PRL.118.123002,Angew.Chem.56.12512}
based on only electric-dipole transition moments had been proposed
with the framework of cyclic three-level systems. Such cyclic three-level systems
can only exist in the chiral molecules and other symmetry broken systems~\cite{SCP,3C,3C1}.

The cyclic three-level systems of chiral molecules are special since the products of the corresponding three Rabi frequencies {can} change sign with enantiomers~\cite{Hirota,Nature.497.475,PRL.111.023008,PCCP.16.11114,ACI,JCP.142.214201,JPCL.6.196,JPCL.7.341,Angew.Chem.10.1002,KK,PRL.87.183002,PRL.90.033001,PRA.77.015403,
PRL.99.130403,JCP.132.194315,JPB.43.185402,PRL.122.173202,PRL.118.123002,Angew.Chem.56.12512}. Accordingly, the two enantiomers will evolve differently with the same initial states.
The inner-state enantio-separation of a chiral mixture is achieved if molecules in one of the three inner states are enantio-pure (i.e., with only one enantiomer occupying that state). The probability of that state among the whole three states for that enantiomer can be defined as the efficiency of the inner-state enantio-separation. The enantio-pure molecules in that state can be further spatially separated from the initial chiral mixture by a variety of energy-dependent processes~\cite{PRL.87.183002,PRL.90.033001}. In the original methods~\cite{PRL.87.183002,PRL.90.033001}, the highly efficient inner-state enantio-separation was realized by means of the concepts from the adiabatic passage techniques~\cite{RMP.89.015006}, which make enantio-separation process~\cite{PRL.87.183002,PRL.90.033001} slow and complicated.

In order to overcome these defects in the adiabatic methods~\cite{PRL.87.183002,PRL.90.033001}, a simple method has been introduced to promote the enantio-separation speed~\cite{PRL.122.173202} by using shortcuts-to-adiabaticity techniques~\cite{OC.139.48,JPCA.107.9937,PRL.105.123003}.
Comparing with the original adiabtical methods~\cite{PRL.87.183002,PRL.90.033001}, the simpler and faster highly efficient enantio-separations can also be achieved by using only dynamic ultrashort-pulse operations~\cite{PRA.77.015403,JPB.43.185402}.
Inspired by the recent breakthrough experiments in enantio-discrimination~\cite{Nature.497.475,PRL.111.023008,PCCP.16.11114,
ACI,JCP.142.214201,JPCL.6.196,JPCL.7.341,Angew.Chem.10.1002} and  separation~\cite{PRL.118.123002,Angew.Chem.56.12512}, some works refocus on the related issues~\cite{PRL.122.223201,OE.27.13965,1807.09425,1905.03956,JCP.149.094201,PRA.98.063401,1904.02208}.

In this paper, we propose two dynamical methods to achieve highly efficient inner-state enantio-separation based on cyclic three-level systems. When the parameters are appropriately adjusted, the evolutions of the two enantiomers initially prepared in their corresponding ground states can be simultaneously described by effective two-level models with the same effective Rabi frequencies but different effective detunings. By further modifying the parameters to ensure that the effective two-level models for the two enantiomers are, respectively, on resonance and in the large-detuning limit, one can achieve the approximately $100\%$-efficiency inner-state enantio-separations when {the probability occupying the ground state of} the enantiomer governed by the effective on-resonance two-level model experiences half-integer periods of its Rabi oscillation. Alternatively, when one of the two-level models for the two enantiomers are on resonance and the other one is detuned (without requiring large detuning), one can obtain exactly $100\%$-efficiency enantio-separations by making {the probabilities occupying the ground states} of the two enantiomers experience half-integer and integer periods of their corresponding Rabi oscillations, simultaneously. After achieving the inner-state enantio-separations by means of the above dynamical methods, the enantio-pure molecules in the ground states can be further spatially separated by a variety of energy-dependent processes~\cite{PRL.87.183002,PRL.90.033001,PRL.122.173202}.

\begin{figure}[h]
  \centering
  \includegraphics[width=0.9\columnwidth]{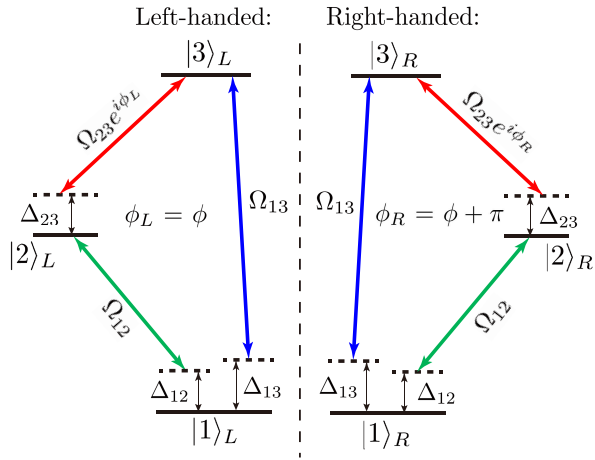}\\
  \caption{Modeling the left- and right-handed enantiomers as cyclic three-level systems where three electromagnetic fields couple respectively to the three electric-dipole transitions. For left-handed enantiomer, the three corresponding Rabi frequencies are $\Omega_{12}$, $\Omega_{13}$, and $\Omega_{23}e^{i\phi}$.
  For the right-handed enantiomer, the three corresponding Rabi frequencies are $\Omega_{12}$, $\Omega_{13}$, and $\Omega_{23}e^{i(\phi+\pi)}$. The detunings for the three transitions are $\Delta_{12}$, $\Delta_{13}$, and $\Delta_{23}$.}\label{Fig0}
\end{figure}

\section{Cyclic three-level systems}\label{ECT}

The two enantiomers can be modeled simultaneously as cyclic three-level systems with choosing appropriate three electromagnetic fields to couple respectively with three electric-dipole transitions~\cite{PRL.87.183002,PRL.90.033001,PRA.77.015403,PRL.99.130403} as shown in Fig.~\ref{Fig0}.
Here, $|j\rangle_{L}$ and $|j\rangle_{R}$ ($j=1,2,3$), which have the same energies $v_{j}$, are the inner states for the left- and right-handed enantiomers, respectively. The scripts $L$ and $R$ have been introduced to denote the left-handed and right-handed enantiomers, respectively. The frequencies of the three electromagnetic fields are $\omega_{12}$, $\omega_{13}$, and $\omega_{23}$, respectively. The detunings of the three transitions are
defined as
\begin{align}
\Delta_{ji}\equiv v_i-v_j-\omega_{ji},~~(3\ge i>j\ge1).
\end{align}

We are interesting in the cases under the three-photon resonance with
\begin{align}
\Delta_{12}+\Delta_{23}=\Delta_{13},
\end{align}
i.e., $\omega_{12}+\omega_{23}=\omega_{13}$.
In the rotating-wave approximation, the cyclic three-level systems for the two enantiomers
can be described in the interaction picture as ($\hbar=1$)~\cite{PRL.87.183002,PRL.90.033001}
\begin{align}\label{HM}
\hat{H}_{Q}=&\Delta_{12}|2\rangle_{QQ}\langle2|+\Delta_{13}|3\rangle_{QQ}\langle3|
+(\Omega_{12}|1\rangle_{QQ}\langle 2|\nonumber\\
&+\Omega_{13}|1\rangle_{QQ}\langle 3|+\Omega_{23}e^{i\phi_{Q}}|2\rangle_{QQ}\langle 3|+h.c.),
\end{align}
where $Q=L,R$ indicate the chirality.
Without loss of generality, we assume $\Omega_{12}$, $\Omega_{13}$ and $\Omega_{23}$ are positive. Here
$\phi_{Q}$ are the overall phases of the three Rabi frequencies for the two enantiomers. The chirality of the cyclic three-level systems is specified by choosing the overall phases of the left- and right-handed enantiomers as
\begin{align}\label{PHM}
\phi_{L}=\phi,~~~~\phi_{R}=\phi+\pi
\end{align}
as shown in Fig.~\ref{Fig0}.

\section{Effective two-level models}

Initially, the two enantiomers are assumed to stay in their ground states $|1\rangle_{L,R}$~\cite{PRL.87.183002,PRL.90.033001,PRA.77.015403,PRL.99.130403,JCP.132.194315,JPB.43.185402,PRL.122.173202}. In the following, we will show that the evolution of the two enantiomers can be simultaneously
described by their corresponding effective two-level models under the conditions
\begin{align}\label{CD}
\Delta_{12}=\Delta_{13}\equiv\Delta,~~\Omega_{12}=\Omega_{13}\equiv\Omega,~~\phi=0.
\end{align}

With the conditions~(\ref{CD}), 
we can rewrite the Hamiltonian~(\ref{HM}) for the two enantiomers as
\begin{align}\label{HM2}
\hat{H}_{Q}=&(\sqrt{2}\Omega|1\rangle_{QQ}\langle D_{+}|+h.c.)+{\Delta}^{Q}_{+}|D_{+}\rangle_{QQ}\langle D_{+}|\nonumber\\
&+{\Delta}^{Q}_{-}|D_{-}\rangle_{QQ}\langle D_{-}|
\end{align}
with
\begin{align}
&{\Delta}^{L}_{\pm}=\Delta\pm\Omega_{23},\nonumber\\
&{\Delta}^{R}_{\pm}=\Delta\mp\Omega_{23},
\end{align}
in the dressed state basis $\{|1\rangle_{Q},|D_{+}\rangle_{Q},|D_{-}\rangle_{Q}\}$ ($Q=L,R$) with
\begin{align}
&|D_{\pm}\rangle_{Q}=\frac{1}{\sqrt{2}}(|2\rangle_{Q}\pm|3\rangle_{Q}).
\end{align}

\begin{figure}[htp]
  \centering
  \includegraphics[width=0.8\columnwidth]{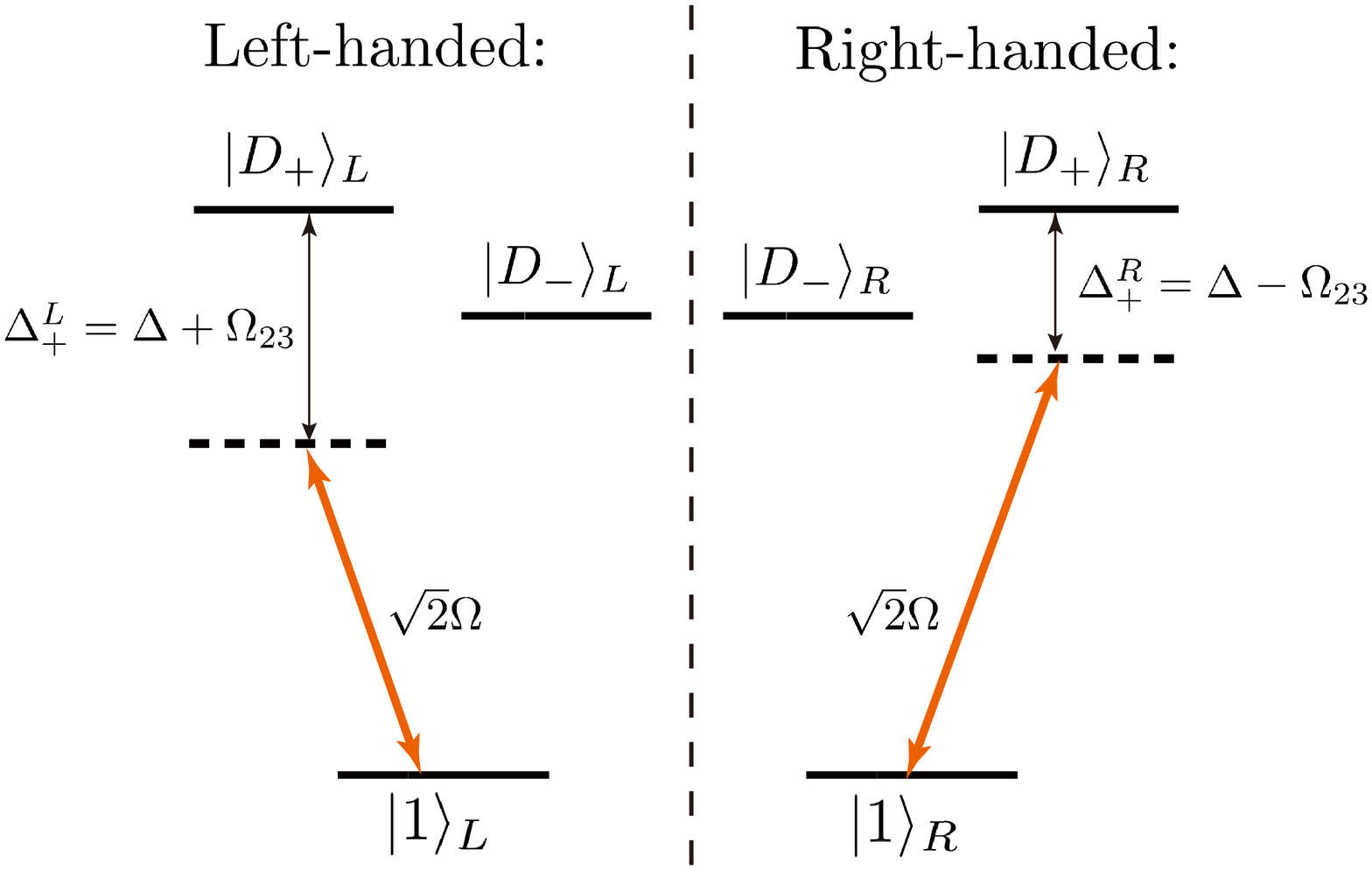}\\
  \caption{Chirality-dependent effective two-level models for the two enantiomers initial prepared in their ground states under the condition (\ref{CD}):  $\Delta_{12}=\Delta_{13}\equiv\Delta$, $\Omega_{12}=\Omega_{12}\equiv\Omega$,  $\phi=0$, and $\Delta_{23}=0$. The dressed states are
  $|D_{+}\rangle_{Q}=(|2\rangle_{Q}+|3\rangle_{Q})/\sqrt{2}$ and $|D_{-}\rangle_{Q}={(|2\rangle_{Q}-|3\rangle_{Q})}/{\sqrt{2}}$. The effective Rabi frequency are $\sqrt{2}\Omega$ and the effective detunings are ${\Delta}^{L}_{+}=\Delta+\Omega_{23}$ and ${\Delta}^{R}_{+}=\Delta-\Omega_{23}$.}\label{Fig1}
\end{figure}

Since $|D_{-}\rangle_{Q}$  decouples with the other states $|1\rangle_{Q}$ and $|D_{+}\rangle_{Q}$, the two enantiomers, initially prepared in their ground states $|1\rangle_{Q}$, will not evolve to the dressed state $|D_{-}\rangle_{Q}$, and can be described by the effective two-level models with
\begin{align}\label{H2}
&\hat{H}^{\mathrm{eff}}_{Q}=(\sqrt{2}\Omega|1\rangle_{QQ}\langle D_{+}|+h.c.)+{\Delta}^{Q}_{+}|D_{+}\rangle_{QQ}\langle D_{+}|.
\end{align}
For the two enantiomers, their corresponding effective Rabi frequencies are $\sqrt{2}\Omega$.
The effective two-level models are chirality-dependent since the effective detunings are different ${\Delta}^{L}_{+}\ne{\Delta}^{R}_{+}$. In Fig.~\ref{Fig1}, we show the cyclic three-level models for the two enantiomers can reduce to the effective two-level ones under the condition (\ref{CD}).

{We would like to note that when $\phi=0$ changes to be $\phi=\pi$ in the condition (\ref{CD}), one will obtain the similar results of effective two-level models except the forms of the Hamiltonian for the two enantiomers exchanging with each other. For convenience, we will focus on the cases with $\phi=0$.}

\section{highly efficient inner-state Enantio-separations}\label{MES}
So far we have shown that the cyclic three-level systems for the enantiomers can reduce to the chirality-dependent effective two-level models.
Further, we will propose two methods to realize highly efficient inner-state enantio-separation with the $0$ and (approximate) $1$ probabilities of the ground state for the
two enantiomers, respectively. {After that, the two enantiomers in different energy states can be spatially separated by a variety of energy-dependent processes~\cite{PRL.87.183002,PRL.90.033001,PRL.122.173202}.}

For the two enantiomers of initial ground states, their evolved states are governed by Eq.~(\ref{H2}) with the corresponding probabilities occupying the ground states
\begin{align}\label{rhoLR}
&P^{Q}_{1}(t)= \frac{4\Omega^2}{\omega^2_{Q}}[\cos(\omega_{Q} t)-1]+1,~~(Q=L,R)
\end{align}
where {the Rabi oscillation frequencies of the probabilities occupying the ground states are}
\begin{align}\label{OMGS}
\omega_{L,R}=\sqrt{8\Omega^2+(\Delta\pm\Omega_{23})^2}.
\end{align}
Here the index $L$ on the left-hand side corresponds to the positive sign on the right-hand side.

{Accordingly, $P^{R}_{1}$ can be equal to $0$ when the effective two-level models for the right-handed enantiomer are on resonance, i.e.,
\begin{align}\label{FCD}
\Delta_{+}^{R}\equiv\Omega_{23}-\Delta=0.
\end{align}{
Then, the Rabi oscillation frequencies of the probabilities occupying the ground states are $\omega_{L}=2\sqrt{2\Omega^2+\Delta^2}$ and $\omega_{R}=2\sqrt{2}\Omega$, and the corresponding Rabi oscillation periods are
\begin{align}
T_{L}=\frac{\pi}{\sqrt{2\Omega^2+\Delta^2}},~~T_{R}=\frac{\sqrt{2}\pi}{2\Omega}
\end{align}}
In what follows, we will show two methods of inner-state enantiomer-separation under the condition~(\ref{FCD}) as well as the condition~(\ref{CD}), by generating the different probabilities occupying the ground states for the two enantionmers.}

\subsection{Approximately $100\%$-efficiency inner-state enantio-separations}

Here we propose the first method to realize the approximately $100\%$-efficiency inner-state enantio-separations with $P^{R}_{1}=0$ and $P^{L}_{1}\simeq1$ based on the conditions~(\ref{CD}) and (\ref{FCD}) and the following one
\begin{align}\label{DOD}
\Delta\gg\Omega .
\end{align}
In this case, the effective two-level models of the two enantiomers are, respectively, in the large-detuning limit and on resonance
\begin{align}\label{CDD1}
{\Delta}^{L}_{+}=2\Delta\gg\sqrt{2}\Omega,~~~~{\Delta}^{R}_{+}=0.
\end{align}
For the left-handed enantiomer, it will always stay approximately in the initial state $|1\rangle_{L}$ due to the large-detuning coupling. On the contrary, the right-handed enantiomer can be totally transferred to the state $|D_{+}\rangle_{R}$
at the time instant {
\begin{align}
t=(n+\frac{1}{2})T_{R}
\end{align}
with the integer $n\geq0$, i.e., when the probability occupying the ground state of the right-handed enantiomer experiences half-integer periods of its Rabi oscillation.} The corresponding probability occupying the ground state for the left-handed enantiomer is {$P^{L}_{1}(t)$ which satisfies $1\ge P^{L}_{1}(t)\ge(1-4\Omega^2/\sqrt{2\Omega^2+\Delta^2})\simeq1$}. Thus the approximately $100\%$-efficiency inner-state enantio-separations is achieved.

\begin{figure}[htp]
  \centering
  \includegraphics[width=0.8\columnwidth]{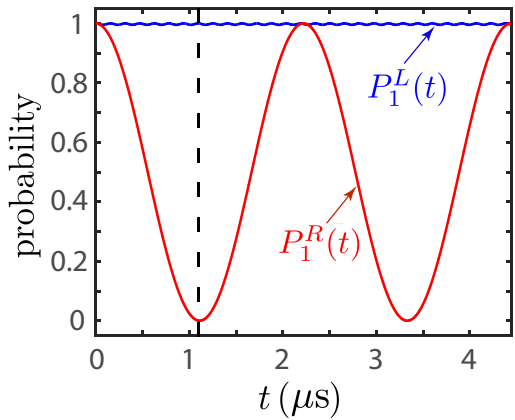}\\
  \caption{Approximately $100\%$-efficiency inner-state enantio-separations. The probabilities occupying the ground states of the left-handed enantiomer $P^{L}_{1}(t)$ (blue line) and the right-handed enantiomer $P^{R}_{1}(t)$ (red line). The parameters are $\phi=0$, $\Delta_{12}=\Delta_{13}\equiv\Delta=20$\,MHz, $\Omega_{23}=20$\,MHz and $\Omega_{12}=\Omega_{13}\equiv\Omega=1$\,MHz. According to the three-photon resonance condition, we have $\Delta_{23}=0$\,MHz. The approximately $100\%$-efficiency inner-state enantio-separation is achieved at e.g. $t=\sqrt{2}\pi/4$\,$\mu \mathrm{s}$ (black dashed line) with $P^{L}_{1}=0.9998\simeq1$ and $P^{R}_{1}=0$.  }\label{Fig3}
\end{figure}

In Fig.~\ref{Fig3}, we numerically demonstrate the approximately $100\%$-efficiency inner-state enantio-separations by referring to the typical experimental parameters~\cite{PRL.118.123002,Angew.Chem.56.12512} as $\Delta_{12}=\Delta_{13}\equiv\Delta=\Omega_{23}=20$\,MHz, $\Omega_{12}=\Omega_{13}=\Omega=1$\,MHz.
We assume the two enantiomers are initially prepared in their ground states~\cite{PRL.87.183002,PRL.90.033001,PRA.77.015403,PRL.99.130403,JCP.132.194315,JPB.43.185402,PRL.122.173202}. As expected, the left-handed enantiomer still stays approximately in the ground state during the evolution.
{The probability occupying the ground state of the right-handed enantiomer experiences a half period of its Rabi oscillation and then becomes $P^{R}_{1}=0$. In the meanwhile, $P^{L}_{1}=0.9998\simeq1$ (see the dashed line in Fig.~\ref{Fig3}).} This clearly indicates a highly efficient inner-state enantio-separation with efficiency of $99.98\%$. This efficiency can be further increased by increasing $\Delta$ and/or decreasing $\Omega$. Similarly, we can also realize the approximately $100\%$-efficiency inner-state enantio-separations with $P^{L}_{1}=0$ and $P^{R}_{1}\simeq1$ {since the evolved probabilities of the two enantiomers occupying the ground state exchange with each other by changing $\Delta$ to its opposite value [seen from Eq.~(\ref{rhoLR})] or replacing $\phi=0$ in Eq.~(\ref{CD}) by $\phi=\pi$.}

\subsection{Exactly $100\%$-efficiency inner-state enantio-separations}
In our first method, the large detuning is needed and the efficiency of inner-state enantiomer-separation is approximate $100\%$. Alternatively, we will propose a method without the requirement of the large detuning to achieve $100\%$-efficiency enantiomer-separation. This would happen if
\begin{align}\label{EHE}
P^{L}_{1}(t)=1, \ \ \ \ P^{R}_{1}(t)=0.
\end{align}
This will happen when the probabilities occupying the ground states of the left- and right-handed enantiomers experience integer and half-integer periods of their corresponding Rabi oscillations, simultaneously. That means it happens at the time instant satisfying
\begin{align}
t=n_{L}T_{L}=(n_{R}+\frac{1}{2})T_{R},~~(n_{L}>n_{R}\ge0)
\end{align}
which requires
\begin{align}\label{CDS}
&\Delta=\frac{\sqrt{8n_{L}^2-2(2n_{R}+1)^2}}{2n_{R}+{1}}\Omega,~~(n_{L}>n_{R}\ge0).
\end{align}

\begin{figure}[htp]
  \centering
  \includegraphics[width=0.8\columnwidth]{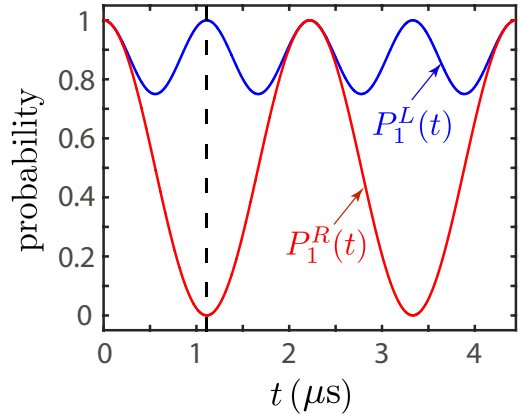}\\
  \caption{Exactly $100\%$-efficiency inner-state enantio-separations. The probabilities occupying the ground states of the left-handed enantiomer $P^{L}_{1}(t)$ (blue line) and the right-handed enantiomer $P^{R}_{1}(t)$ (red line). The parameters are $\phi=0$, $\Delta_{12}=\Delta_{13}\equiv\Delta=\sqrt{6}$\,MHz, $\Omega_{23}=\sqrt{6}$\,MHz $\Omega_{12}=\Omega_{13}\equiv\Omega=1$\,MHz, and $\Delta_{23}=0$\,MHz. The exactly $100\%$-efficiency inner-state enantio-separations is achieved at $t=\sqrt{2}\pi/4$\,$\mu \mathrm{s}$ (black dashed line) with $P^{L}_{1}=1$ and $P^{R}_{1}=0$, when the probabilities of the two enantiomers experiences integer ($1$) and half-integer ($1/2$) periods of their corresponding Rabi oscillations.
}\label{Fig4}
\end{figure}

In Fig.~\ref{Fig4}\,(a), we numerically demonstrate the exactly $100\%$-efficiency inner-state enantio-separations at the typical experimental parameters~\cite{PRL.118.123002,Angew.Chem.56.12512} with $\Delta_{12}=\Delta_{13}\equiv\Delta=\sqrt{6}$\,MHz, $\Omega_{23}=\sqrt{6}$\,MHz, $\Omega_{12}=\Omega_{13}\equiv\Omega=1$\,MHz {under the conditions~(\ref{CD}), (\ref{FCD}), and (\ref{CDS}) with $n_{L}=1$ and $n_{R}=0$.} It clearly shows that the exactly-$100\%$-efficiency inner-state separation with $P^{L}_{1}=1$ and $P^{R}_{1}=0$ simultaneously. {Similarly, we can also realize the exactly $100\%$-efficiency inner-state enantio-separations with $P^{L}_{1}=0$ and $P^{R}_{1}=1$ since the evolutions of the two enantiomers exchange with each other by changing $\Delta$ to its opposite number or choosing $\phi=\pi$.}

\section{Summary}\label{SUM}
We have proposed two methods to achieve highly efficient inner-state enantio-separations based on the cyclic-three level systems of chiral molecules. Then, the enantio-pure molecules in the ground state can be further spatially separated by the energy-dependent processes~\cite{PRL.87.183002,PRL.90.033001,PRL.122.173202}. By choosing the appropriate parameters under the three-photon resonance condition, the two enantiomers initially prepared in their corresponding ground states $|1\rangle_{Q}$ can be governed by chirality-dependent effective two-level models in the basis $\{|1\rangle_{Q},|D_{+}\rangle_{Q}\}$ with $Q=L,R$. Their corresponding effective two-level models have the same effective Rabi frequencies but different effective detunings. The approximately $100\%$-efficiency inner-state separations can be realized when the effective two-level models for the two enantiomers are on-resonance and in the large-detuning limit, respectively. Specifically, the probability occupying the ground state of the enantiomer becomes $0$ after half-integer of its on-resonance Rabi oscillation and in the meanwhile the other enantiomer stays approximately in the ground state.
Moreover, we have proposed the second method to realize exactly $100\%$-efficiency inner-state enantio-separations, when the probabilities occupying the ground states of the two enantiomers experience half-integer and integer periods of their corresponding on-resonance and detuned Rabi oscillations, simultaneously.


Comparing with the original adiabatical enantio-separation methods based on three-level systems of chiral molecules~\cite{PRL.87.183002,PRL.90.033001}, our two methods are faster simpler since {they are based on the dynamical evolutions of the two enantiomers governed by the simple chirality-dependent effective two-level models.} Acutally, our methods are similar to the dynamical enantio-separation methods based on three-level systems~\cite{PRA.77.015403,JPB.43.185402} with reducing the three-level models of the two enantiomers to effective two-level ones. However, they require less steps than the dynamic methods~\cite{PRA.77.015403,JPB.43.185402} to realize the inner-state enantio-separation. Comparing with the shortcuts-to-adiabaticity method~\cite{PRL.122.173202} where the parameters are time-dependently controlled to fulfill the condition of shortcuts to adiabaticity, our methods are simpler since the parameters are time-independent in our proposals.

\section{Acknowledgement}
This work was supported by the National Key R\&D Program of China grant
(2016YFA0301200), the Natural Science
Foundation of China (under Grants No.~11774024, No.~11534002, No.~U1530401, and No.~U1730449),
and the Science Challenge Project (under Grant No.~TZ2018003).

{}

\end{document}